# Gradient Impedance Matching Layers Enable Broadband Water-Air Sound Transmission

Ping Zhou[1,2], Han Jia[2,3*], Yafeng Bi[1], Yunhan Yang[1,2], Yuzhen Yang[1], Peng Zhang[1,2] and Jun Yang[1,2*]

[1] Key Laboratory of Noise and Vibration Research, Institute of Acoustics, Chinese Academy of Sciences, Beijing 100190, People's Republic of China

[2] University of Chinese Academy of Sciences, Beijing 100049, People's Republic of China

[3] State Key Laboratory of Acoustics, Institute of Acoustics, Chinese Academy of Sciences, Beijing 100190, People's Republic of China

[*] Authors to whom correspondence should be addressed: hjia@mail.ioa.ac.cn; jyang@mail.ioa.ac.cn;

## Abstract

Efficient sound transmission across the water-air interface has always been expected in the field of ocean exploration. However, the existing researches are mainly concentrated on the narrow-band transmission based on resonance, which greatly limits the transmission capacity and efficiency. Here, we combined the air-based and water-based metafluids to realize an exponential gradient impedance matching layer for broadband water-air sound transmission. By cooperatively adjusting the sound velocity and thickness in the matching layers, we modulated the required acoustic parameters of each layer into a reasonable range, which can be conveniently achieved by the proposed metafluids. A matching layer sample was constructed and validated in a water tank. Experimental results show that the proposed matching layer can achieve an average sound energy transmission enhancement above 16.7dB from 880Hz to 1760Hz across the water-air interface. Moreover, we use the proposed matching layer to demonstrate a multicolor picture transmission from air to water, which shows extremely high communication capacity and accuracy. Our work is promising for more applications based on water-air transmission and opens a new avenue to the design and implementation of the extreme impedance matching case.

# 1. Introduction

Acoustic communication, which can realize underwater long-range information transmission, has great significance for the ocean exploration applications, such as ocean network development [1-2], ocean geological survey [3-4] and marine life research [5-7]. With the explosive growth of data volume in ocean exploration, there is an urgent need for an efficient transmission method to directly receive underwater information in the air. In Figure 1a, we schematically illustrate the acoustic communication scenario between the underwater and airborne vehicles. When the sound signals emitted by the underwater vehicle impinge on the water-air interface, almost 99.9% of the sound energy is reflected back to the water due to the huge impedance ratio of 3600 between water and air [8], which greatly limits the communication efficiency. Therefore, achieving water-air impedance matching is extremely crucial for efficient water-air acoustic communication.

In recent years, there have been some studies on matching the impedance of water and air. Various types of metasurfaces are demonstrated as impedance transformers between water and air [9-14]. It is reported that the membrane-type metasurface allows about 30% of the incident acoustic power at 700 Hz to be transmitted from water to air [9]. By constructing a mass-spring system, the hydrophobic solid array metasurfaces [10-11] and a bioinspired lotus metasurface [12] are designed to improve the water-air sound transmission at low frequency and intermediate frequency, respectively. Assisted by topology optimization, a metasurface composed of resonance-matching unit cells is designed to realize a water-air sound transmission enhancement of 25.9 dB at the peak frequency [13]. In addition, a composite waveguide with a closed aperture [15] and underwater resonant bubbles close to the interface [16] are also verified as effective ways to enhance water-air sound transmission. However, these researches on water-air impedance matching are all limited to the narrowband transmission based on the resonance, which greatly restricts the transmission capacity. And the broadband impedance matching between water and air still remains a challenge.

Here, we designed a water-air gradient impedance matching layer (GIML) by combining the air-based and water-based metafluids. The water-air GIML is shown schematically in the right panel of Figure 1a. The air-based metafluid is made of square solid inclusions arranged periodically in air and the water-based metafluid is made of square hollow inclusions arranged periodically in water. The realizable effective impedance ranges of the two metafluids can completely cover the gap between air and water. A cooperative design method is proposed to flexibly adjust the sound velocity and thickness of each layer in the water-air GIML, which can modulate the required acoustic parameters of each layer into an achievable range. Utilizing the combination of two metafluids, we designed a four-layer GIML with exponential distribution, which realizes an average sound transmission enhancement over 16.7 dB in the frequency range of 880 Hz-1760 Hz. Based on the broadband transmission ability of the designed GIML, a frequency division multiplexing method is selected to achieve a picture transmission from the air to water with a high capacity. The good performance demonstrates that the proposed GIML has great potential for acoustic communication between air and water.

## 2. Results
### 2.1 Design of the GIML

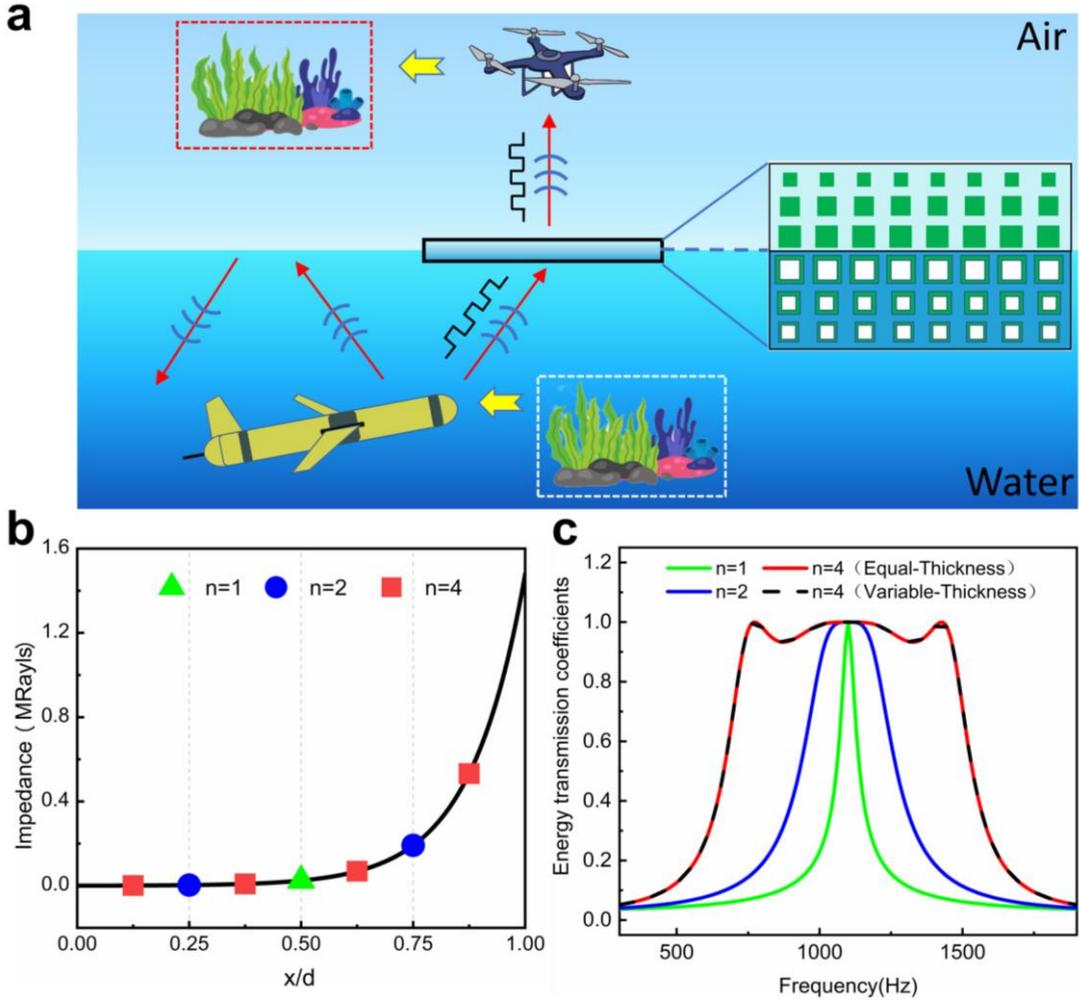

**Figure 1.** The design of the proposed water-air GIML. a) The schematic to show that the proposed GIML can realize the direct acoustic communication between underwater and airborne vehicles. b) The air-to-water exponential impedance distribution (black line) and the discrete impedance value when the water-air impedance matching layer contains 1 layer (green triangle), 2 layers (blue circle) and 4 layers (red square). c) The energy transmission coefficients when the water-air impedance matching layer contains 1 layer (green line), 2 layers (blue line) and 4 layers (red solid line and black dashed line). The red solid line and black dashed line represent the energy transmission coefficients in the equal-thickness and variable-thickness layering methods, respectively. The center frequency is selected as 1100 Hz.

The exponential impedance variation is widely used in the design of the GIML due to its excellent broadband transmission performance [17-20]. Here, we select the exponential impedance variation to couple air and water. The exponential impedance profile from air (at $x = 0$) to water (at $x = d$) is plotted as the black line in Figure 1b, which can be expressed as $Z(x) = Z_a e^{\alpha x}$, where $\alpha = \frac{1}{d}\ln(Z_w/Z_a)$, $Z_a$ is the impedance of air, $Z_w$ is the impedance of water, and $d$ is the thickness of the GIML. Utilizing the equal-thickness layering method, the exponential impedance profile is divided into $n$ layers and the thickness of each layer is $d/n$. Then the impedance of the $i$th $(i = 1,2 \ldots n)$ layer can be expressed as $Z_{i,n} = Z_a e^{\alpha\left(\frac{2i-1}{2n}d\right)}$. Because acoustic impedance depends on both the mass density and the sound velocity, it is necessary to determine the specific material parameters in each layer to achieve the target impedance distribution. As the exponential impedance distribution is the limit case of multi-layer quarter-wavelength matching layers [21], the center transmission frequency $(f_0)$, the thickness of each layer $(d_0 = d/n)$ and the sound velocity in each layer $(c_0)$ should satisfy the relation of $f_0 = c_0/4d_0$ to achieve the high transmission. This means that the sound velocity in each layer should keep the same in the equal-thickness layering method. However, due to the huge impedance difference between water and air, the GIML with a constant sound velocity will lead to a huge mass density difference, which poses great difficulties in the structure realization. To overcome this barrier, we proposed a cooperative design method to flexibly adjust the sound velocity and thickness of each layer in the GIML. With the increasing impedance, we adjust the sound velocity in the $i$th layer($c_{i,n}$) so that the mass density in the $i$th layer($\rho_{i,n}$) can be modulated into a reasonable range. Meanwhile, to guarantee high transmission, we adjust the thickness of the $i$th layer according to $d_{i,n} = c_{i,n}/4f_0$. Then, the total thickness of the GIML is equal to the sum of the thickness of each layer, that is $d = \sum_i d_i$. We take the design of a four-layer GIML as an example. The corresponding impedance distribution is marked as red squares in Figure 1b. The red solid line and black dashed line in Figure 1c show the sound energy transmission coefficients of equal-thickness layering method and variable-thickness layering method, which are completely consistent. The four-layer GIML can achieve energy transmission above 0.9 in the frequency range of 730 Hz to 1470 Hz. The detailed geometric and acoustic parameters in the equal-thickness and variable-thickness layering methods are displayed in **Supplementary Table 1**. We further calculated the sound energy transmission coefficients of the impedance matching layers containing 1 layer and 2 layers, as are plotted in Figure 1c. The corresponding impedance distributions are shown as different symbols in Figure 1b. It can be seen that the transmission bandwidth is expanding with the number of layers. Therefore, we can achieve broadband transmission in a specified bandwidth by choosing an appropriate number of layers. The proposed design method can flexibly adjust the acoustic parameters in each layer of the GIML by changing their geometric parameters, which makes it possible to realize the water-air impedance matching using artificial structures.

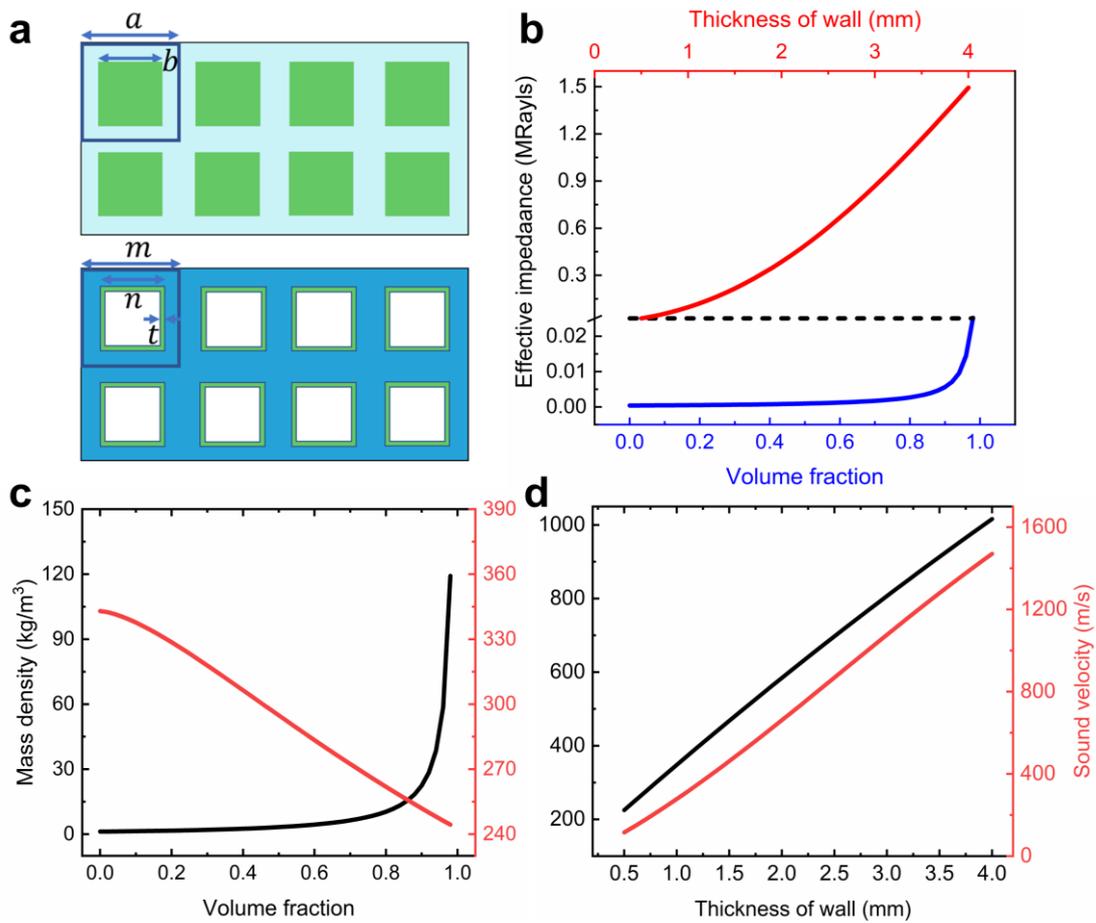

**Figure 2.** The achievable parameters range of the air-based and water-based metafluids. a) The structure schematic of the air-based (top panel) and water-based (bottom panel) metafluids. Both the solid materials are aluminum. b) The achievable effective impedance of the air-based and water-based metafluids. The blue line represents the effective impedance of the air-based metafluid versus volume fraction of square aluminum. The red line represents the effective impedance of the water-based metafluid versus wall thickness of the hollow square frame when $m = 40\ mm$, $n = 38\ mm$. c) The achievable effective mass density (black line) and sound velocity (red line) of the air-based metafluid varying with the volume fraction of square aluminum. d) The achievable effective mass density (black line) and sound velocity (red line) of the water-based metafluid varying with the thickness wall (*t*) of the hollow square frame when *m* = 40 *mm, n* = 38 *mm*.

The effective medium theory is usually used to design metafluids with broadband acoustic parameters [22-29]. The effective mass density ($\rho_e$) and bulk modulus ($K_e$) of the metafluid can be expressed as:

$$\rho_e = \rho_b f_b + \rho_s f_s, \quad \frac{1}{K_e} = \frac{f_b}{K_b} + \frac{f_s}{K_s}. \tag{1}$$

where $\rho$, $K$ and $f$ are the mass density, bulk modulus and the volume fraction, respectively; the subscripts $b$ and $s$ represent the background fluid and the solid inclusions, respectively. Then, the effective impedance and sound velocity of the metafluid can be obtained through $Z_e = \sqrt{\rho_e K_e}, c_e = \sqrt{K_e/\rho_e}$. As can be seen from equation (1), the effective parameters of the metafluid can be adjusted by the volume fraction and material parameters of the components. This enables the widespread application of metafluids in the realization of devices such as gradient-index lens [30-33] and carpet cloak [34-37]. However, only an individual air-based or water-based metafluid is used in these devices, and the effective parameters are always limited near the background medium. It is difficult to realize the huge impedance gradient between air and water by using a single metafluid. Here, we propose a water-air GIML by combining the air-based and water-based metafluids. Two metafluids used to achieve the GIML are shown in Figure 2a. The air-based metafluid is made of square rigid inclusions arranged periodically in the air. The detailed geometric parameters are marked in the figure, wherein $a$ is the lattice constant and $b$ represents the side length of the solid inclusion. We calculated the effective acoustic parameters of the air-based metafluid versus the volume fraction of solid inclusions ($f_s = b^2/a^2$). The calculated sound velocity (red line) and mass density (black line) are shown in Figure 2c. With the increasing volume fraction, the sound velocity of air-based metafluid shows a decrease and mass density of air-based metafluid shows an increase. Accordingly, we can obtain the effective impedance as shown by the blue line in Figure 2b. It can be seen that although the effective impedance of the air-based metafluid varies over a wide range around the air, it is much less than that of water. On the other hand, the water-based metafluid is composed of hollow square inclusions periodically arranged in water. As shown in Figure 2a, $m$ marks the lattice constant, $n$ marks the side length of the solid inclusion and $t$ marks the wall thickness of the hollow square inclusion. We chose aluminum as the solid frame and further calculated the effective acoustic parameters of the water-based metafluid. Figure 2d shows the mass density (black line) and sound velocity (red line) of the water-based metafluids varying with $t$ when $m = 40 mm, n = 38 mm$. Both the mass density and sound velocity of the water-based metafluid increase with $t$. The corresponding effective impedance is shown as the red lines in Figure 2b. It can be seen that although water-based metafluid achieves the acoustic impedance over a wide range, it cannot reach the part close to air. Comparing the red and blue lines in Figure 2b, we can find that there is an overlap between them as marked by the black dashed line. This means that the combination of the two metafluids can exactly fill the impedance gap between air and water, which promises the realization of the water-air GIML. See the calculation process for the effective parameter of the air-based and water-based metafluids in **Supplementary Note 1.**

## 2.2 Experimental demonstration of the GIML

Based on the above analysis, we have designed and demonstrated a water-air GIML by utilizing the two metafluids. The schematic of the designed GIML is shown in the inset of Figure 3a. The GIML contains four layers, whose impedance distribution is the same as the red squares in Figure 1b. The upper two layers are realized by the air-based metafluid, while the lower two layers are realized by the water-based metafluid. The effective sound velocities from the first layer (close to the air) to the fourth layer (close to the water) are 287.9 m/s, 249.0 m/s, 151.2 m/s and 786.8 m/s, respectively. Correspondingly, to keep their center transmission frequencies around 1100 Hz, the thicknesses are set as 60 mm, 60 mm, 36 mm and 180 mm, respectively. The total thickness of the GIML is $d = 336\ mm$. See the detailed geometric and acoustic parameters of the GIML in **Supplementary Table 2**. We calculated the energy transmission coefficients with and without the designed GIML in the finite element package COMSOL Multiphysics. The results are presented in Figure 3b with red and black lines, respectively. Compared with the transmission coefficients of the bare interface close to 0, the designed GIML can achieve the sound energy transmission above 0.7 in the range of 720Hz to 1370Hz, which corresponds to a sound energy enhancement of 28.5 dB.

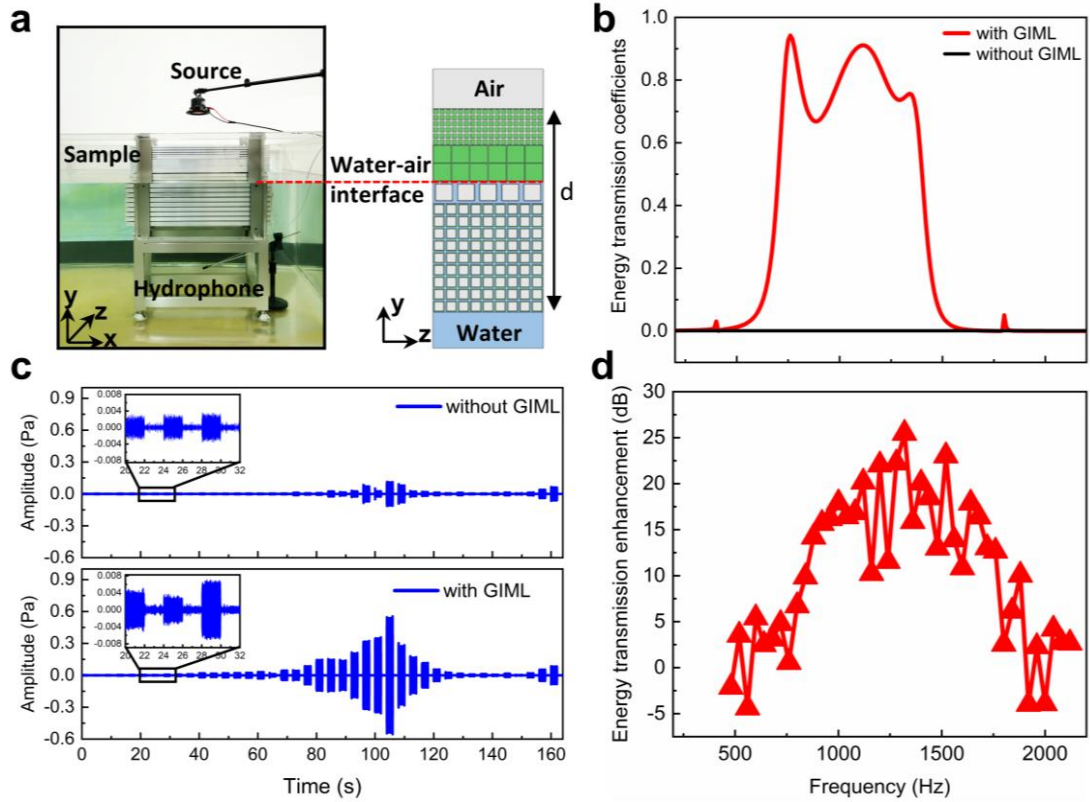

**Figure 3.** The experimental demonstration of the GIML. a) Photograph of the experimental setup for the transmission enhancement measurement of the GIML. Inset: a schematic of the geometric profile of the GIML. The upper two layers close to air are air-based metafluid and the lower two layers close to water are water-based metafluid. b) The simulated results of the energy transmission coefficients with (red line) and without (black line) the GIML. c) The received acoustic pressure signal without (top panel) and with (bottom panel) the GIML. Insets: enlarged view of the acoustic pressure signals from 20 s to 32 s. d) The experimental result of the ETE with the GIML.

Then, we carried out a qualitative measurement to verify the energy transmission enhancement (ETE) of the GIML. The experimental setup is shown in Figure 3a. The aluminum pipes with required geometric sizes were inserted into two identical perforated plates to form a quasi-2D GIML sample. A loudspeaker was mounted above the sample in the air as a source. A hydrophone was fixed below the sample in the water tank to record the transmitted signal. In the measurement, the loudspeaker emitted sine bursts from 480 Hz to 2120 Hz with an interval of 40 Hz. The duration of each frequency signal is 2 seconds, and the pause between the adjacent frequency signals is 2 seconds. The hydrophone recorded signals without and with the GIML, respectively. The received time-domain pressure signals are shown in Figure 3c. The top and bottom panel shows the received time signal without and with the GIML, respectively. It is obvious that the overall pressure amplitude is significantly enhanced after installing the GIML. Due to the frequency response characteristic of the loudspeaker, the received pressure amplitudes are relatively weak at some frequencies. To show these signals more clearly, an enlarged view of the time signals from 20s to 32s in two cases are shown in the insets of Figure 3c, which show that the pressure amplitudes at these frequencies are all significantly higher than the background noise. To further analyze the performance of the GIML, a fast Fourier transform is performed on the time-domain signals. Then we extracted the pressure amplitudes in different frequencies and calculated the ETE of the GIML versus the frequency as shown in Figure 3d. The result shows that the average ETE of the GIML is over 16.7 dB in the frequency range of 880 Hz – 1760 Hz. Comparing the simulation with experimental results, they both demonstrated the broadband transmission enhancement effect of the proposed GIML. There exists a slight deviation in ETE and frequency band, which originates from the open measurement environment and the fabrication error in sample size.

## 2.3 Acoustic communication based on the GIML

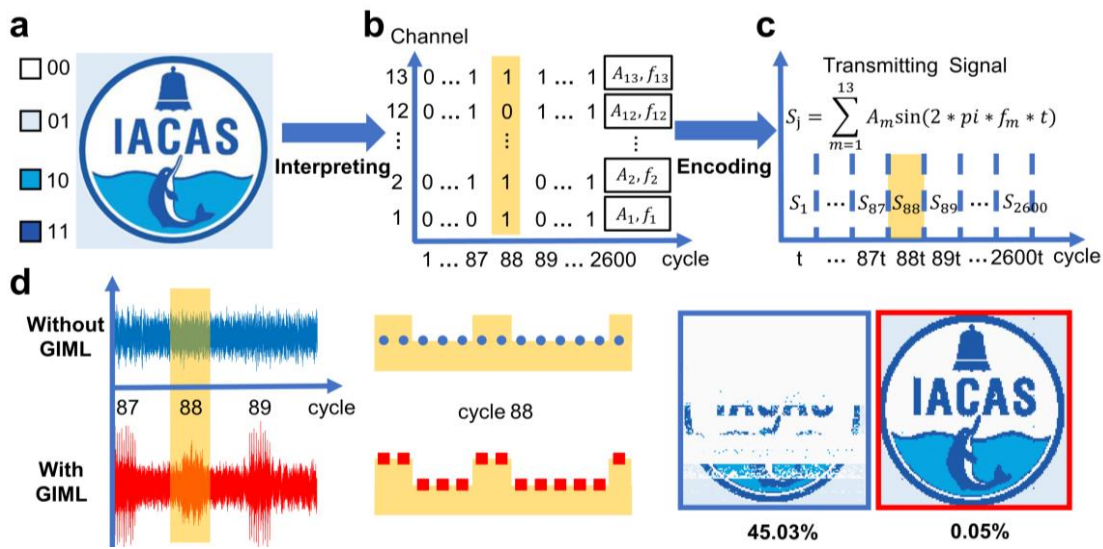

**Figure 4.** Experimental demonstration of the water-air acoustic communication. a) The transmitted

picture including four different colors and 130*130 pixels. Each pixel is converted into a 2-bit code based on different colors. b) The transmitted bit stream allocated to 2600 cycles and 13 channels. According to the frequency-division multiplexing method, 13 frequencies ($f_m$) are selected to carry the bit values ($A_m$). c) The time-domain signal $S_j = \sum_{m=1}^{13} A_m * \sin(2 * pi * f_m * t)$ encoded by the bit stream. d) Left panel: the received time-domain signals in 87th-89th cycle without (blue line) and with (red line) GIML. Middle panel: comparison between the target code (orange ladder) and the decoded results for the 88th cycle without (blue circles) and with (red squares) the GIML. Right panel: the received complete pictures without (blue frame) and with (red frame) the GIML. The percentages (45.03% and 0.05%) below the pictures represent the bit error ratio in two cases.

With the broadband transmission enhancement ability between air and water, the proposed GIML is promising for applications in acoustic transmission between air and water. Here, we use the designed GIML to realize the air-to-water acoustic communication. In this part, we aim to transmit the picture in Figure 4a from air to water by frequency division multiplexing. The target picture has a size of 130*130 pixels and contains 4 different colors. By denoting the navy blue as '11', the cyan as '10', the light blue as '01' and the white as '00', the target picture is converted into a bitstream with a size of 130*260 bits. The schematics of bitstream allocation and encoding are shown in Figures 4b and 4c, respectively. The bitstream is allocated to 2600 cycles. Each cycle contains 13 channels and is encoded to a time-domain signal, which is expressed as $S_j = \sum_{m=1}^{13} A_m * \sin(2 * pi * f_m * t), (j = 1,2 \dots 2600, m = 1,2,\dots,13)$, where $A_m$ denotes the bit value of each channel, $f_m$ denotes the frequency assigned to each channel and $t = 0.3s$. The data capacity of each cycle is 13 bits, then an aggregate capacity of 42.9 bits/s is obtained. Here, the frequencies are selected based on the ETE bandwidth of the designed GIML, which are expressed as $f_n = 1060 + (n-1) * 40$. A time-domain sequence containing 2600 coded signals is emitted from the air and received in the water. The received signals were decoded by marking the target signal with a signal-to-noise ratio (SNR) over 20dB as "1" otherwise marking it as "0". Finally, the experiment was conducted in the tank. The experimental setup is the same as that in Figure 3a. We emitted and measured the signals with and without the GIML, respectively. The received transmission signals of the 87th-89th cycle are extracted and shown in the left panel of Figure 4d. The decoding result of the 88th cycle is shown in the middle panel of Figure 4d, where the orange ladders mark the target code and the symbols mark the decoding results. From the comparison, the received coded signals without the GIML are almost submerged in the background noise, and the decoding results are all '0'. After inserting the GIML, the SNR of the received coded signals is significantly enhanced and the decoding result is completely consistent with the target code. The received complete pictures in two cases are displayed in blue and red frames in the right panel of Figure 4d, respectively. It is obvious that the received picture through the bare interface can hardly be recognized with a bit error rate up to 45.03%, while the received picture through GIML is almost the same as the target picture with a negligible bit error rate of 0.05%. Thus, the broadband impedance matching between air and water can promote the applications in acoustic communication between water and air.

## 3. Conclusion

In conclusion, we combine air-based and water-based metafluids to design a water-air GIML. Both the simulated and experimental results demonstrate the good performance of broadband sound transmission enhancement between water and air. Utilizing the designed GIML, we also demonstrated the air-to-water acoustic communication by transmitting a picture from air to water. The proposed water-air GIML in this work opens a new avenue for both design and applications of future acoustic metamaterials. On the one hand, the reported cooperative design method and material system provide a simple, effective and low-cost way for broadband impedance matching between extreme media like water and air, which is promising for more applications such as nondestructive testing and biomedical ultrasonography. On the other hand, the broadband high transmission property of the designed water-air GIML not only endows a high communication capacity, but also achieves a negligible bit error rate, which makes it appealing for numerous ocean exploration applications requiring effective acoustic communication between water and air. Last but not least, the low-frequency high transmission property of the GIML shows great potential for application in underwater noise reduction based on ocean ecological protection (See the detailed verification in **Supplementary Note 2**.). Our work will greatly facilitate the development of new broadband impedance matching devices and ocean acoustic communication.

## 4. Experimental Section

*Sample fabrication:* All the unit cells of the designed metafluids are fabricated into aluminum pipes with a length of 600 mm. Both ends of the unit cells of water-based metafluids are sealed with glass cement to prevent water from penetrating. Then, all these pipes are inserted in two identical perforated plate supports for fixing. Finally, the sample is fixed on a bracket and placed in an organic glass water tank.

*Experimental measurement:* In the measurement of ETE, the experiment is performed in an organic glass water tank with a size of 1.5m*1m*0.75m. We choose a loudspeaker of M4N, HiVi as the sound source and a hydrophone of Type 8130, B&K as the underwater detector. All the emitting and receiving acoustic signals are analyzed by a multi-analyzer system (Type 3560, B&K). The ETE is obtained from the difference of the acoustic energy transmission with and without the GIML. It is expressed as $ETE = 20\log(p_t/p_{t0})$ where $p_t$ and $p_{t0}$ are the measured underwater pressure amplitudes with and without the GIML, respectively.

*Numerical simulation:* The numerical simulations are performed with the acoustic-solid interaction module in COMSOL Multiphysics. In the simulations of Figure 3b, the material parameters of water are set as: $\rho$=1000 kg/m$^3$, K=2.19 GPa; the material parameters of air are set as: $\rho$=1.21 kg/m$^3$, c= 343 m/s; the material parameters of aluminum are set as: $\rho$= 2700 kg/m$^3$, E=70 GPa, $\mu$=0.33; The designed GIML is placed between air and water region. The right and left ends of all the regions are set as rigid boundaries, and all the interfaces between solid materials and fluid (air and water) are set to be acoustic-structure boundaries. A uniform plane wave background field is applied in the air domain to provide an incident pressure field. The plane wave radiation is applied in the bottom boundary of the water region. The calculated frequencies vary from 200 Hz to 2200Hz with an interval of 10 Hz.

## Supporting Information

Supporting Information is available from the Wiley Online Library or from the author.

## Acknowledgements

This work is supported by the Key-Area Research and Development Program of Guangdong Province (Grant No. 2020B010190002), the National Natural Science Foundation of China (Grant No. 11874383, 12104480), the IACAS Frontier Exploration Project (Grant No. QYTS202110).

## Conflict of Interest

The authors declare no conflict of interest. Keywords hydrogels, metamaterials, tunable acoustics

## Keywords

broadband water-air sound transmission, metafluid, cooperative design